\documentclass{svproc}

\usepackage[colorinlistoftodos]{todonotes}

\usepackage{color} 
\PassOptionsToPackage{hyphens}{url}
\usepackage{hyperref}
\hypersetup{colorlinks=true,
	urlcolor=blue,citecolor=blue}

\usepackage{amsmath}  
\usepackage{amssymb}
\usepackage{amsfonts}
\usepackage[hmargin=2cm,vmargin=3cm]{geometry}
\usepackage{graphicx}
\graphicspath{ {images/} }
\usepackage[utf8]{inputenc}
\usepackage[T1]{fontenc}
\usepackage{eurosym}

\usepackage{mdframed}

\usepackage{cite}

\begin{document}
\title{The Road to 
Quantum 
Computational Supremacy}
	 \titlerunning{The Road to 
Quantum 
Computational Supremacy}
\author{Cristian S. Calude\inst{1} \and Elena Calude\inst{2}}

\institute{Department of Computer Science, University of Auckland\\
Private Bag 92019, Auckland, New Zealand\\ \email{c.calude@auckland.ac.nz}
\and Institute of Natural and {\color{black}Computational} Sciences,
Massey University at Albany\\
Private Bag 102-904 North Shore MSC, 
Auckland, New Zealand \\ \email{e.calude@massey.ac.nz}}


\maketitle

 
\begin{abstract}
We present an idiosyncratic view of the race for quantum computational supremacy.
Google's approach and IBM challenge are examined.  An unexpected side-effect of the race is the
significant progress in designing fast classical algorithms. Quantum supremacy, if achieved, won't make classical computing obsolete.

\end{abstract}	

\begin{flushright}
{\it 
A hyper-fast quantum computer is the digital equivalent of\\
a nuclear bomb; whoever possesses one will be able to\\ shred 
any encryption and break any code in existence.\footnote{A typical example of incorrect, largely-spread,  myth quoted from a recent mystery novel.}}~\cite{Ignatus2018}

\end{flushright}

\section{Fairy tales or more cautionary tales?}

Following the development of  Shor's quantum algorithm~\cite{shor:94} in 1994 and Grover's quantum algorithm~\cite{grover} two years later, quantum computing was seen as a bright beacon in computer science, which led to a surge of theoretical and experimental  results. The field captured the interest and imagination
of the large public and media, and not surprisingly, unfounded claims about the power of quantum computing and its applications proliferated.

 A certain degree of pessimism began to infiltrate  when experimental groups floundered while attempting to control more than a handful of qubits.   Recently, 
 a broad wave of ambitious industry-led research programmes in quantum computing---driven by D-Wave Systems,\footnote{The company's relatively steady progress  in producing and selling  the first series of D-Wave quantum computers has gone from 28 qubits in 2007 to more than 2,000 in their 2000Q$^{\rm TM}$ System machine~\cite{dwavesys2017}. {\color{black} In September 2019 the  5,000-qubit D-Wave machine called ``Advantage'' has been delivered  to the Los Alamos National Laboratory~\cite{Advantage2019}.}}
  the  tech giants Google, IBM, Microsoft, Intel and startups like Rigetti Computing and Quantum Circuits Incorporated---has emerged\footnote{Of course, the industry work is based and has continued the
  academic efforts, sometimes using successful experimentalists from academia, like Google {\color{black}does.}} 
  and  bold claims about a future revolutionised by quantum computing are resurfacing. 

\begin{figure}[ht]
\begin{center}
		\includegraphics[scale=.58]{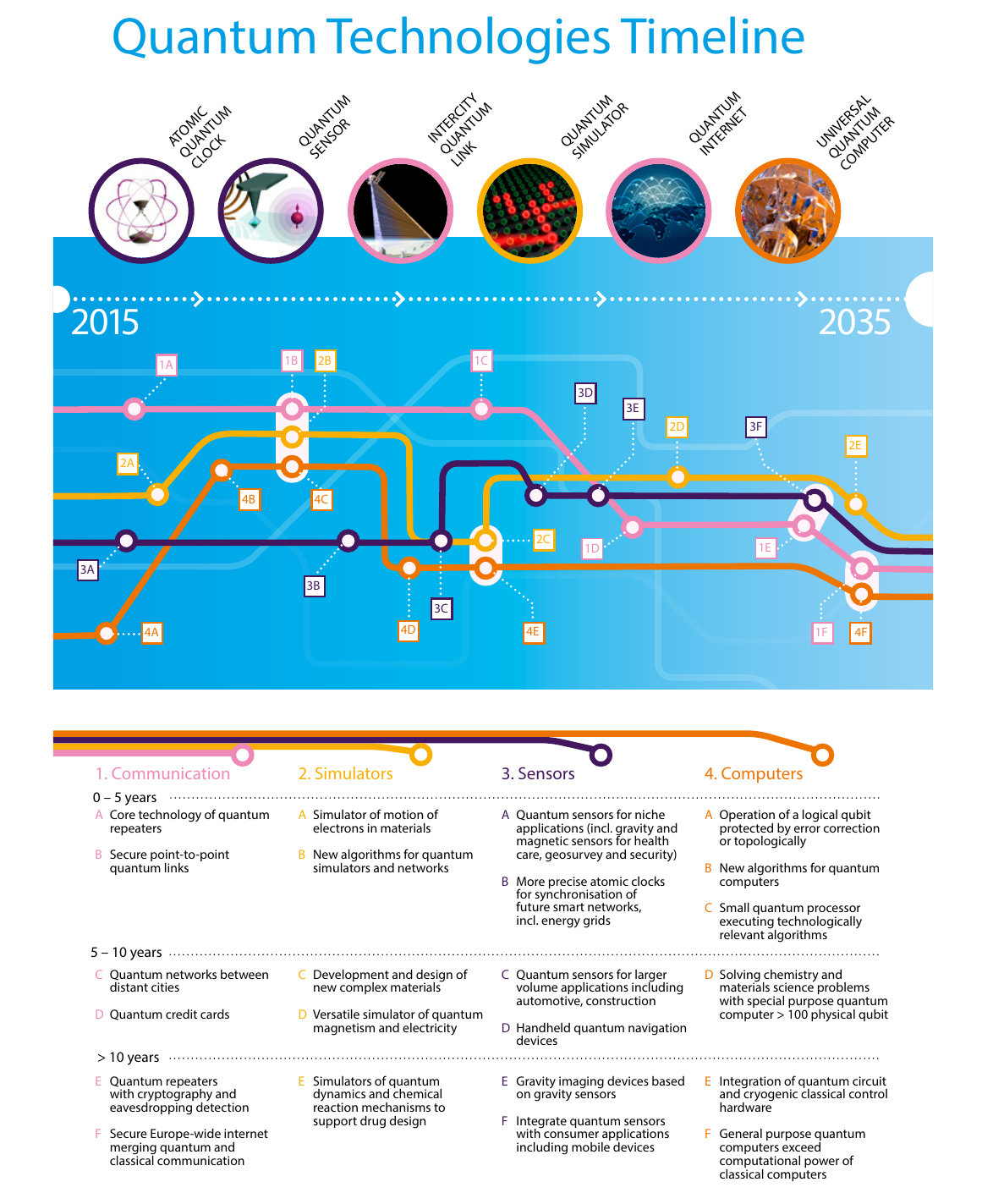}
		\label{EUQflagship}
\caption{Quantum timeline: 2015--2035, \cite{Qmanifesto2016}}
\end{center}
\end{figure}

 Governments are also  involved:   phase 1 (2015--2019) \textsterling 330 million  of the UK government  programme on quantum technologies~\cite{QCUK} is rolling and the European Commission has announced a \euro 1 billion  initiative in  quantum technology~\cite{QCEU2017}.
The European flagship quantum programme, whose explicit goal is to stimulate a ``second quantum revolution'', aims to ``build a universal quantum computer able to demonstrate the resolution of a problem that, with current techniques on a supercomputer, would take longer than the age of the universe'' by 2035, \cite{Qmanifesto2016}; see also Figure~\ref{EUQflagship}.

Undoubtably, these programmes are extremely beneficial to the development of various quantum technologies,  but,  are the claims about the future of quantum computing realistic?  ``We tend to be too optimistic about the short run, too pessimistic about the long run'' said recently J.~Preskill~\cite{preskill_oct2017}; see also~\cite{AACCQblog2016,svozil-2016-quantum-hokus-pokus}.

\section{Quantum algorithmics}
\label{qalgorithmics}

{\color{black} First and foremost, {\em quantum computing cannot compute all partial functions  a universal Turing machine can calculate because only total functions can be computed by quantum circuits}~\cite{quassical_ned2018}.
Consequently, quantum computing potential advantages could come only from  faster than classical computations.}

 While Shor's algorithm, Deutsch-Jozsa algorithm and various others in the ``black-box'' paradigm\footnote{Where access to a quantum black-box or ``oracle'' with certain structural properties is assumed.} are believed to
 provide an exponential speedup over classical computers, this is far from the case in general. We said ``believed'' because the superiority of Shor's quantum algorithm over classical ones is still an open problem and various techniques allowing efficient classical simulation of quantum algorithms have been successfully developed~\cite{Griffiths:1996aa,Browne:2007aa,Abbott:2012ix}
 even for some ``black-box'' quantum ones~\cite{Calude:2007aa,Abbott:2011aa,Johansson2017,Johansson:2019aa}.

In fact, since the introduction of Shor's and Grover's algorithms some twenty years ago, the development within the field of quantum algorithmics has been rather 
slow---see~\cite{quantum_alg} for a global picture---and many of them  are novel uses of a handful of core quantum algorithms.  So, why are there  so few quantum algorithms  that offer speed-up over classical algorithms? Although written more than a decade ago, 
Shor's article~\cite{Shor:2003:WHM:602382.602408} is still actual:

\begin{quote}
	The first possible reason is that quantum computers operate in a manner so different from classical computers that our techniques for designing algorithms and our intuitions for understanding the process of computation no longer work. The second reason is that there really might be relatively few problems for which quantum computers can offer a substantial speed-up over classical computers, and we may have already discovered many or all of the important techniques for constructing quantum algorithms.
\end{quote}

Best quantum algorithms  typically provide a quadratic or low-order polynomial speedup~\cite{Furer:2008aa}. Furthermore, there are pointers~\cite{be-va,limitsQC-2008} suggesting that quantum computers  
cannot offer more than a (perhaps small) polynomial advantage for  
{\bf NP}-complete problems,\footnote{Perhaps the most important class of ``difficult computational problems'' such as the well known travelling-salesman problem, which have applications in almost every area of science and beyond, from planning and logistics to microchip manufacturing.} and such a speedup would struggle to compete with the heuristic approaches commonly used to solve them in practice. 
 However, even  a polynomial-order speedup could be of significant benefit for problems requiring exact solutions or for problems that can classically be solved in sub-exponential time, like the graph isomorphism problem (see~\cite{graphisoqubo2019}).

Grover's quantum algorithm~\cite{grover} is an interesting example:   access to an unsorted quantum database that can be queried with a quantum input  is given, and asked if it contains a specific entry.  Grover's algorithm offers a {\em provable} speedup. However,  the speedup is not exponential and, more importantly, the problem it solves is far from being realistic: the cost of constructing the quantum database could negate any advantage of the algorithm, and in many classical scenarios one could do much better by simply creating (and maintaining) an ordered database.  Using  Grover's algorithm  as a subroutine  for solving problems in image processing  is more efficient because the cost of preparing the quantum ``database'' can be spread out over several calls~\cite{Lanzagorta:2005:HQC:1198555.1198723}; this strategy motivated a new hybrid quantum-classical paradigm for embedded quantum annealing algorithms~\cite{doi:10.1142/S0219749919500424}. Other applications are discussed in~\cite{QuantAlgorithms2016}.

Quantum simulation, quantum-assisted optimisation and quantum sampling are believed to
offer near-term quantum solutions to hard problems that may lead even  to commercialisation~\cite{qcommer2017}.

\section{What is quantum 
computational supremacy?}
\label{wQS}
The quantum computational advantage for simulating quantum systems
was first stated by Feynman in 1982, in one of the pioneering papers in quantum computing~\cite{feynman} (the other one was   Manin~\cite{Manin_1980}).   What is the justification of Feynman's insight?  According to the data processing inequality~\cite{informationtheory91,Beaudry:2012:IPD:2230996.2231000},   (classical)  post-processing cannot increase information.  This suggests that to run an accurate classical simulation of a quantum system  one must know a lot about the system before the simulation is started~\cite{allen_email_nov2017}.  Manin~\cite{Manin_1980} and Feynman~\cite{feynman} have argued that a quantum computer might not need to have so much knowledge. 
This line of reasoning   seemingly inspired  Deutsch~\cite{Deutsch:1985aa}  to state
\begin{quote}
	{\bf The postulate of quantum computation}: Computational devices based on quantum mechanics will be computationally superior compared to digital computers.
\end{quote}
A spectacular support for this postulate came from Shor's 1994 polynomial factoring quantum algorithm~\cite{shor:94} in spite of the fact that the problem whether factoring is in {\bf P} was, and 
still is,  open. The belief that factoring integers is computationally hard\footnote{For results pointing to the opposite assumption
see~\cite{Griffiths:1996aa,Browne:2007aa,Abbott:2012ix,Nene2016,post_quatumRSA}.}  is essential  for 
 much of modern cryptography and computing security.     In 2002  Hemaspaandra,  Hemaspaandra and Zimand~\cite{HEMASPAANDRA2002171}, improving results in~\cite{Simon_1997,berth_brassard1992}, showed that  there are tasks on which polynomial-time quantum machines are exponentially faster almost everywhere than any classical -- even bounded-error probabilistic -- machine.

In 2011 the syntagm ``quantum supremacy'' was coined and discussed\footnote{The use of  the word ``supremacy''---which denotes ``the state or condition of being superior to all others in authority''---was criticised in~\cite{qsupremacyname2017} because  the syntagm  `white supremacy' is associated with 
the racial segregation and discrimination of the apartheid regime of South Africa.
Proposals like ``quantum advantage'' or ``quantum superiority''   have been discussed~\cite{qsupremacyblog2017}, but to date none has gained ground.}  by J.~Preskill  in his Rapporteur talk  ``Quantum Entanglement and
Quantum Computing''~\cite{jpreskill:2012} at the 25th {\em Solvay Conference on Physics} (Brussels, Belgium, 19--22 October
2011):

\begin{quote}
	We therefore hope to hasten the onset of the era of quantum supremacy, when we will be able to perform tasks with controlled quantum systems going beyond what can be achieved with ordinary digital computers. 
\end{quote}

Recently, quantum supremacy was described  in~\cite{charact_quantum_sup2017} as follows:

\begin{quotation}
Quantum supremacy is achieved when a formal computational task is performed with an existing quantum device which cannot be performed using any known algorithm running on an existing classical supercomputer in a reasonable amount of time.
\end{quotation}

Note the imprecision in the above  formulation: the comparison is made with ``any known algorithm running on an existing classical supercomputer'' and the  classical computation takes ``a reasonable amount of time''. Can this imprecision be decreased  or, even better, eliminated?
Just as there is no current proof that {\bf P} $\not=$ {\bf NP}---one of the important open problems in classical complexity theory---there is no mathematical proof for the Postulate of quantum computation; in fact, the Postulate is not  amenable to a proof. The hypothesis {\bf P} $\not=$ {\bf NP}  can be used for deriving useful results; similarly, 
adopting assumptions in  terms of  both quantum physics and classical complexity theory---which can be justified 
heuristically or experimentally---can lead to precise statements which can be proved or disproved.  The following  two assumptions 

\begin{quote}
{\bf The postulate of noise}: Quantum systems are inherently noisy.	
\end{quote}

\begin{quote}
	{\bf The Extended Church-Turing Thesis}: 
	A probabilistic Turing machine can efficiently simulate any realistic model of computation.
\end{quote}

have been used by Kalai~\cite{Anti_quantum_supremacy2011} to challenge the Postulate of quantum computation. Here ``efficiently'' means ``with at most polynomial overhead'';  the adjective ``realistic'' (or  ``reasonable'' as an alternative) refers to a ``physically realisable in principle''.
 It is worth mentioning that these assumptions are themselves challengeable; see for example~\cite{charact_quantum_sup2017} for
the Extended Church-Turing Thesis.

A quantum computational supremacy experiment has to prove both a lower bound and an upper bound.
  In Google's proposed experiment---to be discussed in details in Section~\ref{google_sup}---the upper bound is given by a
quantum algorithm running  on a quantum computer with 49 qubits\footnote{A qubit is a 2-state quantum system. There are many ways to build qubits, hence   not all qubits are equal. The magic number 49 (or 50) refers to qubits in the quantum circuit model which are more difficult to control than the qubits used by the D-Wave machine~\cite{calude-elena-dinneen15} (to embed a complete graph of $N$ vertices in D-Wave hardware Chimera graph we need approximately  $N^2$  qubits, so 2,048 D-Wave qubits correspond to about  fully connected 45 qubits) or the trapped atom qubits used by specialised quantum simulators~\cite{Bernien:2017aa,Zhang:2017aa}.}---a mathematical fact and an engineering artefact  (the construction of the quantum machine);  the lower bound is 
necessary for proving that no current classical computer can simulate the sampling in a reasonable time from the output distributions of pseudo-random quantum circuits. 

Upper bounds are positive results while lower bounds are negative. Upper bounds are useful when we want to show that a problem can be  solved by a ``good''  algorithm. But if we want to argue that  no algorithm solving a  problem can be better than a given one, or perhaps that some problem is so hard that we can't possibly hope to find a good solution to it, we need lower bounds.

 In mathematics and theoretical computer science it is  well-known  that negative results are more difficult to prove than positive ones.  In classical computability theory it is more difficult to prove incomputability than computability, and in complexity theory lower bounds are more difficult to prove than  upper bounds~\cite{Sipser:1996:ITC:524279}.  The superiority of Shor's quantum algorithm~\cite{shor:94}  is a prime example. A methodology for proving lower bounds in quantum computing is discussed in~\cite[p.~144--149]{Gruska}.
   Sometimes unproved claims about the quantum superiority of a quantum algorithm have been shown to be incorrect: an example is the superiority of Deutsch's quantum algorithm over any classical one, see~{\color{black}\cite{Deutsch:1985aa,Calude:2007aa,Johansson2017}}. 
   
   Another issue is correctness: how do we know that the quantum computer solution is indeed correct---quantum computing is a probabilistic type of computation---if we can't check it with a reliably tested classical computer? 
{\color{black}   For a promising approach see~\cite{aharonov2017interactive,howtoverifyQC}.}
   Meantime we note that even classical correctness is a very difficult problem. The Ackermann $A$ function~\cite{ackermann:28}  is a singular example: computing the value
   of $A(x,y)$ is prohibitively difficult because the function is computable but not primitive recursive,
   but testing the predicate $A(x,y)=z$ is very easy~\cite{DBLP:journals/ipl/Calude87}. 

Finally, the discussion about quantum supremacy  suggests 
 a misleading comparison between classical and quantum computing. If a quantum computer can outdo {\bf any} classical computer on one problem  we have quantum supremacy, even if classical computers could be at least as good as quantum ones in solving  many (most) 
 other problems.

 Put it bluntly, {\em quantum supremacy, if achieved, won't make classical computing obsolete.}  In fact, the hybrid approach combining quantum  and classical computing, briefly mentioned in Section~\ref{qalgorithmics},  could be a good strategy in solving some (many) difficult problems~\cite{doi:10.1142/S0219749919500424}.

\section{Criteria for quantum computational supremacy}
\label{criteria}

Harrow and Montanaro~\cite{Harrow:2017aa} have  proposed a  reasonable
list of criteria for a quantum supremacy experiment. According to them we need to have:

\begin{enumerate}
	\item a well-defined computational problem,
	\item a quantum algorithm solving  the problem which can run on a near-term hardware capable of dealing with noise and imperfections,
	\item an amount of computational resources (time/space) allowed to any classical competitor,
	\item  a small number of well-justified  complexity-theoretic  assumptions,
	\item a verification method that can efficiently distinguish between the performances of the quantum algorithm from {\bf any} classical competitor using the allowed resources.
\end{enumerate}

Large integer factoring is a  typical problem for  a quantum supremacy experiment. Indeed, it is well-defined, it has huge practical importance, there are efficient quantum algorithms solving it (Shor's algorithm and variants~{\color{black}\cite{shor:94,post_quatumRSA})},  the complexity-theoretic assumption is that no classical algorithm can factor essentially faster than the current ones and the solution is quickly verifiable. This seems an almost ideal candidate, except for a) the strong complexity-theoretic assumption~\cite{Nene2016} and b)   the lack of a near-term hardware  running such a quantum algorithm for sufficiently large integers (say a 2,048-bit number), see~\cite{Harrow:2017aa}. A possible solution for b) could be a hybrid  (quassical) approach~\cite{doi:10.1142/S0219749919500424}.

Harrow and Montanaro~\cite{Harrow:2017aa} state that ``we do not require that the computational task\footnote{Their formulation for what we call a computational problem.} is of practical interest''. This is a strong assumption in itself which is adequate only for a foundational study. 

Table~1 in~\cite{Harrow:2017aa}, p.~205,  lists seven plausible approaches to quantum computational supremacy:
factoring, single photons passing through a linear-optical network  (boson sampling),
quantum circuits on many qubits and only a few
layers of quantum gates (low-depth circuits),  random quantum circuits containing gates that either all commute  or do not commute (instantaneous quantum polynomial-time, IQP), quantum approximate optimisation algorithms (QAOA),  quantum adiabatic optimisation and quantum analogue simulation. These approaches are then evaluated according to usefulness,  assumption implying no classical simulation and difficulties to solve on a quantum computer and to verify. Factoring is the only useful problem, simulation is often useful, adiabatic optimisation could be useful and the remaining three problems do not seem to be useful. Factoring is the hardest to solve on a quantum computer, boson sampling, adiabatic optimisation and analogue simulation are easy and the remaining three are moderately difficult. Only factoring
 is easy to verify. The complexity-theoretic assumptions are generally very strong, assessing their plausibility is a very difficult task and, generally, conclusions are rather controversial. 
 A detailed complexity-theoretic analysis of various possible quantum supremacy experiments  can be found in~\cite{aaraonsonQS2016}.
 The papers~\cite{aaraonsonQS2016,Harrow:2017aa} are exceptionally singular in offering balanced  and more formal  analyses.

\section{Is the quest for quantum computational supremacy worthwhile?}

Apart publicity and marketing, is the effort of demonstrating the quantum computational supremacy 
justified? What are the (possible) benefits? Can the claim of quantum computational supremacy  be falsified?

We will start  with the second question. The main benefit could be foundational and philosophical:  a better understanding of the nature of quantum mechanics through its computational capabilities.\footnote{A  beautiful result regarding the computational power of algorithmic random strings was proved in~\cite{ch-schw-78}. This was used  as a test of quality for quantum randomness in~\cite{PhysRevA.82.022102}.} Such a gain will  boost the efforts of not only building larger-scale quantum computers but also, and,  more importantly, developing new and powerful algorithms for these machines possibly  leading to solutions to important practical problems.  From this perspective the answer to the first question is affirmative.

Let us examine closer  the foundational gain.
A successful quantum supremacy experiment could be a complement to Bell experiment: the latter refuted local hidden models of quantum mechanics, while the former {\em seems} to  invalidate the  Extended Church-Turing Thesis~\cite{yao-2003}. 
 The paper~\cite{Harrow:2017aa}  discusses the advantages of a successful quantum supremacy experiment, even one that barely surpasses any classical competitor, illustrated  with hard-to-simulate classical systems like protein folding or fluid dynamics. Here we suggest a different perspective which motivated the tentative
 formulation above.
The Extended Church-Turing Thesis---which incidentally has nothing to do with either Church nor Turing---is a foundational principle of classical complexity theory which ensures that the  polynomial time class  {\bf P}  is well defined.\footnote{The Thesis equating feasible computation with polynomial-time computation has significantly less ``evidence'' than the Church-Turing Thesis;  in fact,  according to~\cite{davis-interview}, it ``lacks evidence''.} The Thesis places strong constraints, one of them being that  {\em  the model of computation is digital}. For example, analog computers are excluded because they assume infinite arithmetic precision. Furthermore, it is known that an infinite precision calculator with operations +, x, =0?, can factor integers in polynomial time (see~\cite{shamir_factoring1979,Tamma2016}).\footnote{Feyman's 1982 intuition (Section~\ref{wQS}) was  substantiated in~\cite{uqs1996} by  running a quantum analog emulation. The quantum version of analogue computers, continuous-variable quantum computers, have been theoretically studied~\cite{Kendon3609}; the model in~\cite{PhysRevLett.119.120504} offers a universal gate set  for both qubits and continuous variables.}  But,   are quantum computers a ``reasonable'' model of computation?  Are quantum systems digital? At first glance quantum computers (and, more generally, quantum systems) appear to be analog devices, since a quantum gate is described by a unitary transformation, specified by complex numbers;  a more in-depth analysis is still required.

What does it take to refute the claim of quantum computational supremacy?  This amounts to prove that any computation performed by any quantum computer can be simulated by a classical machine in polynomial time, a weaker form of the  Extended Church-Turing Thesis. This statement cannot be proved for the same reasons the Church-Turing Thesis cannot be proved: obviously, they may be disproved.  The paper~\cite{classicalbosonsampling2017}  presents efficient classical boson sampling algorithms and a theoretical analysis of the possibility of scaling boson sampling experiments; it concludes  that ``near-term quantum supremacy via boson sampling is unlikely''.

\section{Google quantum computational supremacy}
\label{google_sup}
In the landscape of various proposals for quantum computational supremacy
experiments  Google's approach is not only  well documented, but  had chances to be completed  really very soon~\cite{google_quantum_supremacy2017}. The  proposed experiment is not about solving a problem: it is the 
computational task of sampling from the output distribution of 
pseudo-random quantum circuits built from a universal gate set.\footnote{For another promising quantum simulation see~\cite{random_compilerQS2019}.}
This computational task is difficult because as the grid size increases, the  {\em memory needed to store everything increases classically exponentially}.\footnote{But, do we {\em really need}  to store everything?}  The required memory
for a $6 \times 4=24$--qubit grid is just 268 megabytes, less than  the average smartphone, but  for a $6 \times 7=42$--qubit grid it jumps to  70 terabytes, roughly 10,000 times that of a high-end PC. Google has used Edison, a supercomputer housed by the US National Energy Research Scientific Computing Center and ranked 72 in the Top500 List~\cite{EdisonJune2017}, to simulate the behaviour of the grid of 42 qubits.
The classical simulation stopped at this stage because going to the next size up {\em was thought to be currently impossible: a $48$-qubit grid would require 2,252 petabytes of memory, almost double that of the top supercomputer in the world. }  The path to quantum computational supremacy was obvious: 
 if Google could solve the problem with a 50--qubit quantum computer, it would have beaten every other computer in existence.

The abstract of the main paper describing the theory behind the experiment~\cite{charact_quantum_sup2017}  reads:\footnote{Our emphasis.}

\begin{quote}
A critical question for the field of quantum computing in the near future is whether quantum devices without error correction can perform a well-defined computational task beyond the capabilities of state-of-the-art classical computers, achieving so-called quantum supremacy. {\em We study the task of sampling from the output distributions of (pseudo-)random quantum circuits, a natural task for benchmarking quantum computers. Crucially, sampling this distribution classically requires a direct numerical simulation of the circuit, with computational cost exponential in the number of qubits.} This requirement is typical of chaotic systems. {\em We extend previous results in computational complexity to argue more formally that this sampling task must take exponential time in a classical computer. } We study the convergence to the chaotic regime using extensive supercomputer simulations, modeling circuits with up to 42 qubits---the largest quantum circuits simulated to date for a computational task that approaches quantum supremacy. We argue that while chaotic states are extremely sensitive to errors, quantum supremacy can be achieved in the near-term with approximately fifty superconducting qubits. We introduce cross entropy as a useful benchmark of quantum circuits which approximates the circuit fidelity. We show that the cross entropy can be efficiently measured when circuit simulations are available. {\em Beyond the classically tractable regime, the cross entropy can be extrapolated and compared with theoretical estimates of circuit fidelity to define a practical quantum supremacy test.}
\end{quote}

Google was on track to deliver  before the end of the year.  Alan Ho, an engineer in Google's quantum AI lab,
 revealed the company's progress at a quantum computing conference in Munich, Germany. According to~\cite{Google_breakthrough2017}:
 
\begin{quote}
His team is currently working with a 20--qubit system that has a ``two--qubit fidelity'' of 99.5 per cent---a measure of how 
error-prone the processor is, with a higher rating equating to fewer errors.	For quantum supremacy, Google will need to build a 
49--qubit system with a two--qubit fidelity of at least 99.7 per cent. Ho is confident his team will deliver this system by the end of this year. 
\end{quote}

Let us note that many, if not most, discussions about quantum computational supremacy focus on the most exciting possibilities of quantum computers, namely the upper bound. What about the lower bound? The article~\cite{charact_quantum_sup2017} refers cautiously to the lower bound in the abstract:
``We extend previous results in computational complexity {\em to argue more formally} that this sampling task must take exponential time in a classical computer.''
	Indeed, they   do not claim to have a proof
for the lower bound, just a ``better formal argument''. Their argument is    reinforced later   in the introduction:

\begin{quote}
	State-of-the-art supercomputers cannot simulate universal random circuits of sufficient depth in a 2D lattice of approximately $7 \times 7$ qubits with any known algorithm and significant fidelity.
\end{quote}

Does Google's experiment satisfy the criteria discussed in Section~\ref{criteria}? The  problem is  well-defined, albeit a simulation, not  a  computational problem,\footnote{One could argue that   the task itself is rather uninteresting and without obvious applications. Indeed,  all the time nature is doing quantum `things' that we don't know how to solve classically. For example, the structure of atoms can in general only be determined experimentally, but nature manages it with near perfect fidelity.  If Google  achieved the goal---an un-disputable big technical feat---the meaning of the achieved ``supremacy'' could still be debatable.} the quantum algorithm solving  the problem will  run on  a quantum computer---promised to be built before the end of 2017\footnote{ ``When pressed for an update, a spokesperson [for Google] recently said that `we hope to announce results as soon as we can, but we're going through all the detailed work to ensure we have a solid result before we announce'.~\cite{era_quantum2018}, 24 January 2018. {\color{black}The goal was not reached as of 30 September 2019.}}---capable of dealing with noise and imperfections, the 
classical competitor would be allowed a reasonable amount of computational resources and there is a plausible verification.  The weakest part comes from the  complexity-theoretic  assumption~\cite{charact_quantum_sup2017}:

	\begin{quote}
{\bf Memory assumption}.  Sampling this distribution classically requires a direct numerical simulation of the circuit, with computational cost exponential in the number of qubits.
\end{quote}

The assumption was corroborated by the statement:
	
	\begin{quote}
		Storing the state of a 46--qubit system takes nearly a petabyte of memory and is at the limit of the most powerful 
		computers.~\cite{google_quantum_supremacy2017}
	\end{quote}

\section{IBM challenge}
\label{ibmchallenge}
The Memory assumption  is crucial  for  the proposed lower bound, and, indeed, this was confirmed very soon.
The paper~\cite{IBMclassical_simu2017}  proved that a supercomputer can simulate sampling from random circuits  with low depth (layers of gates) of up to 56 qubits.

%

\begin{quote}
	With the current rate of progress in quantum computing technologies, 50--qubit systems will soon become a reality. To assess, refine and advance the design and control of these devices, one needs a means to test and evaluate their fidelity. This in turn requires the capability of computing ideal quantum state amplitudes for devices of such sizes and larger. In this study, we present a new approach for this task that significantly extends the boundaries of what can be classically computed. We demonstrate our method by presenting results obtained from a calculation of the complete set of output amplitudes of a universal random circuit with depth 27 in a 2D lattice of $7\times 7$ qubits. We further present results obtained by calculating an arbitrarily selected slice of 237 amplitudes of a universal random circuit with depth 23 in a 2D lattice of $8\times 7$ qubits. Such calculations were previously thought to be impossible due to impracticable memory requirements. {\em 
	Using the methods presented in this paper, the above simulations required 4.5 and 3.0 TB of memory, respectively, to store calculations, which is well within the limits of existing classical computers.\footnote{{\color{black}Our emphasis.}}}
\end{quote}

Better results have been quickly announced, see for example~\cite{quantum_simulationQC2018}.
The limits of classical simulation are   not only  unknown, but hard to predict.

In spite of this, IBM has announced  a prototype of a 50--qubit quantum computer,  stating  that it  ``aims to demonstrate capabilities beyond today's classical systems" with quantum systems of this size~\cite{IBM50Qubit2017}.

\section{Latest developments}

At  2018 Consumer Electronics Show in Las Vegas,  Intel CEO Brian Krzanich reported ``the successful design, fabrication and delivery of a 49-qubit superconducting quantum test chip''~\cite{intel49_2018}.  The 49-qubit superconducting quantum test chip is called  ``Tangle Lake'' after a chain of lakes in Alaska known for extreme cold temperatures. At the event,
Mike Mayberry, managing director of Intel Labs said:  ``We expect it will be five to seven years before the industry gets to tackling engineering-scale problems, and it will likely require 1 million or more qubits to achieve commercial relevance.'' In~\cite{arXiv:1801.00862} J.~Preskill aptly said: ``Quantum computers with 50-100 qubits may be able to perform tasks which surpass the capabilities of today's classical digital computers, but noise in quantum gates will limit the size of quantum circuits that can be executed reliably. \dots Quantum technologists should continue to strive for more accurate quantum gates and, eventually, fully fault-tolerant quantum computing.''  Jay Gambetta, from  IBM Thomas J. Watson Research Center  believes that ``a universal fault-tolerant quantum computer, which has to use logical qubits, is still a long way off'', \cite{era_quantum2018}.   E.~Tang (then an 18-year-old  undergraduate student at UT Austin)  has recently proved~\cite{Tang:2019:QCA:3313276.3316310} that classical computers 
can solve  the ``recommendation problem'' -- given incomplete data on user preferences for products, can one quickly and correctly predict which other products a user will prefer? -- with performance  comparable to that of a quantum computer.  Is this significant?  {\bf Yes}, because  
 quantum computer scientists had considered this problem  to be one of the best examples of a problem that quantum computers can solve exponentially faster 
than their classical ones and the quantum solution in~\cite{Qrecommendation_system2017} was hailed as one of the first examples in {\em quantum machine learning and big data}  that would be unlikely to be done classically\dots  \  {\color{black}In October 2018 Bravyi, Gosset and K\"{o}ning~\cite{shallow_circuits2018} have presented an argument -- based on non-locality -- which suggests that a certain quantum algorithm requiring only constant-depth quantum circuits  can be a suitable candidate for showing quantum computational supremacy. }

\section{Closing remarks}
{\color{black}Recall that the computational power of quantum computing  is less than that of  a universal Turing machine~\cite{quassical_ned2018}, so 
quantum computing potential advantages could come only from  faster than classical computations.}

Does the paper~\cite{IBMclassical_simu2017} destroy the quest for quantum computational supremacy?
 Is there any incompatibility between the classical simulation reported in~\cite{IBMclassical_simu2017} and the IBM statement cited at the end of Section~\ref{ibmchallenge}? Tentatively we answer with  no  to both questions. 
The following paragraph~\cite{aaronson_ibm2017} is relevant:

\begin{quote}
This paper\footnote{That is, \cite{IBMclassical_simu2017}.} does not undercut the rationale for quantum supremacy experiments. The truth, ironically, is almost the opposite: it being possible to simulate 49--qubit circuits using a classical computer is
a precondition for Google's planned quantum supremacy experiment, because it's the only way we know to check such an experiment's results! The goal, with sampling-based quantum supremacy, was always to target the  ``sweet spot,''  which we estimated at around 50 qubits, where classical simulation is still possible, but it's clearly orders of magnitude more expensive than doing the experiment itself. If you like, the goal is to get as far as you can up the mountain of exponentiality, conditioned on people still being able to see you from the base. Why? Because you can. Because it's there.\footnote{``It is not the mountain we conquer but ourselves'',  as  Edmund Hillary aptly said. } Because it challenges those who think quantum computing will never scale: explain this, punks! But there's no point unless you can verify the result.	\end{quote}

Here are a few  more lessons. The first is not  to underestimate 
 the importance of mathematical modelling and  proving (lower bounds, in particular). As the title of the  blog~\cite{aaronson_ibm2017} says, ``$2^n$ is exponential, but $2^{50}$ is finite'', the difference between exponential and polynomial running times is asymptotic and in some concrete cases it  is a challenge
  to find finite evidence for the difference.  Furthermore, proving that a problem is in  {\bf P} itself  is not  a guarantee that there is an algorithm in {\bf P}  that is practically useful:  
   primality has been known to be  in {\bf P} since 2002, but all known deterministic algorithms are  too slow in practice,  so probabilistic tests of primality continue to be used.
 
 Secondly, the conversation on quantum computing, quantum cryptography and their applications  needs an infusion of modesty (if not humility), more technical understanding and clarity as well as less hype. Raising false expectations could be harmful for the  field.
 
 Thirdly, a trend in quantum computing is emerging: when a problem  is solved efficiently in quantum computing, it draws more attention and often produces better classical alternatives than existed before.  Some of the new efficient classical solutions, see for example~\cite{Calude:2007aa,Abbott:2011aa,Abbott:2010ab,Johansson2017,Tang:2019:QCA:3313276.3316310,Johansson:2019aa},  
have been directly inspired by the quantum work. 
 
 Finally, the race quantum vs.~classical is running so fast---a sample is given by  the references posted/published  since October 2017, the month when the paper~\cite{IBMclassical_simu2017} was posted---that by the time this paper  is printed some   results discussed here could be obsolete.  One fact is certain: as of 30 September 2019, the quantum computational supremacy was not (yet?) demonstrated.
 {\color{black}
 \section{P.S. Quantum desperation\protect\footnote{Added on \today.}}
 
  Two important articles have been published in quantum computing on 23-24 October 2019. The first, written by a Google team and  published in the prestigious journal {\em Nature}~\cite{Arute2019}, announces  the experimental realisation of quantum supremacy  with a programmable machine with 53 qubits:
 \begin{quote}
%
 	 {\it Our Sycamore processor takes about 200 seconds to sample one instance of a quantum circuit a million times our benchmarks currently indicate that the equivalent task for a state-of-the-art classical supercomputer would take approximately 10,000 years. This dramatic increase in speed compared to all known classical algorithms is an experimental realization of quantum supremacy for this specific computational task, heralding a much-anticipated computing paradigm.}
 \end{quote}

 This paper has sparked a huge interest not only in the quantum community, but the whole world. Announcements and comments have instantly appeared  in prestigious science magazines like  {\em New Scientist}, 
``Google reigns supreme" and ``It's official: Google has achieved quantum supremacy",  major newspapers like
{\em The Washington Post}, 
``Bravo for Google's `quantum supremacy.' Here's what needs to happen next"
 and worldwide broadcasters like  {\em BBC}, ``Google claims `quantum supremacy' 
 for computer".  Not everybody was convinced even at an intuitive level of understanding:  {\em Reuters}:
 ``Google unveils quantum computer breakthrough; critics say wait a qubit'',  {\em The Financial Post}: ``Google claims `quantum supremacy' with quantum computer breakthrough, but skeptics don't agree'', to cite only two sources. 
 
 The inventor of the concept of quantum supremacy is also cautious~\cite{Preskill_2019}:
 
  \begin{quote}
The Google team has apparently demonstrated that it's now possible to build a quantum machine that's large enough and accurate enough to solve a problem we could not solve before\dots 
\end{quote}

 The second paper, written by an IBM team,  was posted in the archive~\cite{ibm_qs_0ct2019} and is summarised as follows:

\begin{quote} We argue that an ideal simulation of the same task can be performed on a classical system in 2.5 days and with far greater fidelity. This is in fact a conservative, worst-case estimate, and we expect that with additional refinements the classical cost of the simulation can be further reduced. 
\end{quote}
 
 Could both  reports be correct?  
 
 Interestingly,  immediately after publishing the paper~\cite{Arute2019}  {\em Nature} published also an anonymous editorial~\cite{editorialQSNature2019} including the following significant paragraphs:

\begin{quote}As the world digests this achievement  --  including the claim that some quantum computational tasks are beyond supercomputers --  it is too early to say whether supremacy represents a new dawn for information technology. \dots \
At the very least, quantum computers as a routine part of life are likely to be decades or more into the future.

 \dots \medskip
 
Instead of proceeding with caution, a quantum gold rush is under way, with investors joining governments and companies to pour large sums of money into developing quantum technologies. Unrealistic expectations are being fuelled that powerful general-purpose quantum computers could soon be on the horizon. Such misguided optimism could be dangerous for the future of this still-fledgling field.
\end{quote}

Undoubtedly Google's technological achievement is remarkable and it helps  building the case for a possible quantum supremacy by achieving a high upper bound. {\em The real problem is that there is no  formal argument for the lower bound,} see Section~\ref{criteria}, the supplementary material to~\cite{Arute2019} and the mathematical discussion in~\cite{QS2019}. Furthermore, 
{\em it is not for IBM\footnote{Although they did before~\cite{IBMclassical_simu2017}.} or anybody else to disprove the lower bound claimed by Google\footnote{Comments like ``Tellingly, not even IBM thinks the simulation would be especially easy -- nor, as of this writing, has IBM actually carried it out.'' ~\cite{saaronsonNYT2019}  are irrelevant.}: the onus is on Google to prove it.}

Where does ``desperation''  in the title of this section come from? As noted in~\cite{heavenQS2019}

\begin{quote}
	It has taken Google  13 years\footnote{And  a tone of money. (our comment)} 
  to get this far.	
	Without  a profitable device,   research could dry up. It happened to Apollo, programme. It has happened  at times with AI.
\end{quote}

There are very few agreements in quantum computing, but one is that the area has been showered with money in recent years but has delivered very little practical solutions. How long can the flow of money continue? There is a sense that the answer is not too encouraging, so something had/has to be done. Downgrading the mathematical notion of quantum speedup\footnote{Note that Grover's quantum algorithm~\cite{grover} {\em proved a quantum speedup} 25 years ago, yet insufficient to justify quantum computing practicality.} to quantum supremacy was meant to help, but  not without a price. The origin, merit and pitfalls of this concept have been recently discussed by its inventor J.~Preskill in a thoughtful article in {\em Quanta Magazine}~\cite{Preskill_2019}. One main objection pointed there is that the ``word exacerbates the already overhyped reporting on the status of quantum technology". This was echoed also in the IBM paper~ \cite{ibm_qs_0ct2019}:

 \begin{quote}
	For the reasons stated above, and since we already have ample evidence that the term ``quantum supremacy'' is being broadly misinterpreted and causing ever growing amounts of confusion, we urge the community to treat claims that, for the first time, a quantum computer did something that a classical computer cannot with a large dose of skepticism due to the complicated nature of benchmarking an appropriate metric.
\end{quote}

The current tendency seems to move the arguments from the mathematics and science to media propaganda.

\begin{quote}
Google's demonstration should give these skeptics pause. To all appearances, a 53-qubit device really was able to harness 9 quadrillion amplitudes for computation, surpassing (albeit for a special, useless task) all the supercomputers on earth. Quantum mechanics worked: an outcome that's at once expected and mind-boggling, conservative and radical.~\cite{saaronsonNYT2019} 
\end{quote}

Interestingly, this is not a new tactic   in settling quantum mechanics controversies and a most prominent example  is the famous Einstein-Bohr disagreement on the Copenhagen interpretation.  Einstein view~\cite[p. 29]{AdamBecker2018}:

\begin{quote}
  The theory reminds me a little of the system of delusions of an exceedingly intelligent paranoiac. 
 \end{quote}

opposed Bohr's ``shut up and calculate!'' attitude (using Mermin's expression~\cite{mermin_pillow}). According to Lakatos~\cite[p. 59--60]{lakatos_1978},\cite[p. 105]{Lakatos1976}:

 \begin{quote}
  After 1925 Bohr and his associates introduced a new and unprecedented lowering of critical standards for scientific theories. This led to a defeat of reason within modern physics and to an anarchist cult of incomprehensible chaos.  \end{quote}
 
Recently there is an apparent change~\cite[p. 9]{defeat_reason2018}:

\begin{quote}
 while Einstein won and would continue to win all the logical battles, Bohr was decisively winning the propaganda war. 	  
 \end{quote}
 
 Let's hope that in  quantum computing
 mathematics and science will prevail over propaganda.
}

\section*{Dedication} This paper is dedicated to the memory of Jon Borwein (1951--2016) whose broad mathematical interests included also quantum computing.

\section*{Acknowledgment}
We thank  N.~Allen for fruitful discussions and suggestions, specifically for insight on Feynman's paper~\cite{feynman}, and R.~Brent, R.~Goyal, L.~Hemaspaandra,  K.~Pudenz, R.~Hua, K.~Svozil and an anonymous referee for excellent critical comments and suggestions.
This work has  been supported in part by the  Quantum Computing Research Initiatives at Lockheed Martin. 
 
\if01 
\bibliography{cris_s}{}
\bibliographystyle{abbrv}
\fi

\end{document}